\documentclass[structabstract]{aa}

\usepackage{txfonts}
\usepackage{graphicx}

\begin{document}
\newcommand{\chg}{}
\newcommand{\hhh}{\mbox{H$_2$ }}

\newcommand{\ma}{\mbox{m\AA}}
\newcommand{\ms}{\mbox{ms$^{-1}$}}
\newcommand{\kms}{\mbox{kms$^{-1}$}}

\title{Robust limit on a varying proton-to-electron mass ratio from a single
{\mbox{H$_2$}} system}

\author{
M. \,Wendt
\inst{1} 
\and P. \, Molaro\inst{2}
          }

\institute{Hamburger Sternwarte, Universit\"at Hamburg, Gojenbergsweg 112,
21029 Hamburg, Germany\\
\email{mwendt@hs.uni-hamburg.de}
\and
Osservatorio Astronomico di Trieste, Via G.\,B.\,Tiepolo 11,
34131 Trieste, Italy}


\date{Received \today; accepted -- --}

\abstract
{The variation of the dimensionless fundamental physical constant  $\mu=m_p/m_e$
can be checked
 through observation of Lyman and Werner lines of molecular hydrogen in
the spectra of
 distant QSOs. Only few, at present four, systems have been used for the  purpose
providing
 different results  between the different authors. }
{Our intention is to asses the accuracy of the investigation concerning a 
possible variation of the fundamental physical constant $\mu=m_p/m_e$ and to
provide more robust results.
The goal   in mind is to resolve the current controversy on variation of $\mu$
and
devise
explanations for the different findings. We achieve this not by another single
result but
by providing alternative approaches to the problem.}
{The demand for precision requires a deep understanding of the errors involved. 
Self-consistency in data analysis and effective techniques to handle unknown
systematic errors are essential.
An analysis based on independent data sets of QSO 0347-383 is put forward and
new approaches for some
of the steps involved in the data analysis are introduced. In this work we
analyse two independent sets of observations of the same absorption system and 
for the first time we apply corrections for the observed offsets between discrete spectra {\chg mainly caused by slit illumination effects}.}
{Drawing on two independent observations of a single absorption system in
QSO 0347-383 our detailed analysis
yields $\Delta\mu/\mu = (15 \pm (9_{\mathrm{stat}} + 6_{\mathrm{sys}})) \times
10^{-6}$ at $z_{\mathrm{abs}}=3.025$.
Based on the overall 
goodness-of-fit we estimate the limit of accuracy to $\approx$ 300 \ms,
consisting of roughly 180 \ms due to
the uncertainty of the fit and about 120 \ms allocated to systematics.}
{Current analyses tend to underestimate the impact of systematic errors. This
work presents
alternative approaches to handle systematics and introduces methods required
for precision 
analysis of QSO spectra available in the near future.}


\maketitle

\section{Introduction}

 The Standard Model of particle physics  (SMPP) is   very successful and its
predictions are
tested to high  precision in laboratories around the world.  SMPP  
needs   several dimensionless
fundamental constants, such as coupling constants and mass ratios, whose values
cannot be predicted and   must be established through 
experiment (Fritzsch \cite{Fritzsch09}).     Our confidence in their constancy
stems from 
laboratory experiments 
over human time-scales but  variations might have occurred over the 14
billion-year history of the Universe
while  remaining  undetectably small today. 
Indeed, in  theoretical models    seeking to unify
the four forces of nature, the coupling constants vary naturally on cosmological
scales. 
 
The proton-to-electron mass ratio, $\mu  = m_p / m_e$  has been the
subject of numerous studies. The mass ratio  is  sensitive primarily   to
the quantum chromodynamic scale.  The $\Lambda_{QCD}$  scale
should vary considerably faster than that of quantum electrodynamics
$\Lambda_{QED}$.
As a consequence, the secular change in the proton-to-electron mass ratio, if
any, should be larger than that of the fine structure constant.   
This makes   $\mu$  a very interesting target to search for
possible cosmological variations of the fundamental constants.

The present value of the proton-to-electron
mass ratio is  $\mu$ = 1836.15267261(85) (Mohr et al. \cite{Mohr00}). Laboratory
experiments  by comparing  the rates between clocks based on hyperfine
transitions
in atoms with   a different dependence on $\mu$    restrict the
 time-dependence of  $\mu$     at 
the level of {$ (\dot{\mu}/ \mu)_{t_0}  = (1.6 \pm 1.7) \cdot  10^{-15}$
yr$^{-1}$} (Blatt et al. \cite{Blatt08}).

 A probe  of the variation of $\mu$   is obtained by comparing 
rotational
versus vibrational modes of molecules as first suggested by  Thompson
(\cite{Thompson75}).  The
method is based on the fact that the wavelengths of vibro-rotational
lines of molecules depend on the reduced mass, M, of the molecule.   The energy
difference between  two consecutive levels of the rotational
spectrum of a diatomic molecule scales with the reduced mass M, whereas 
the energy difference between two adjacent levels of the vibrational spectrum is
proportional to $(M)^{1/2}$:
 \begin{equation}
   \nu =  c_e + \frac{c_{v}}{   \mu^{1/2}}  +\frac{c_{r}}{ \mu},
 \end{equation}  
{\chg with $c_e$, $c_v$ and $c_r$ as constant factors for the electronic,
vibrational and rotational contribution, respectively.}
Consequently,
by studying the Lyman and Werner transitions of molecular hydrogen we may obtain
 information about  a change in $\mu$.  
The observed wavelength $\lambda_{\mathrm{obs}}$  of any given
line in an absorption system at the redshift $z$ differs from the local rest-frame
wavelength $\lambda_0$  of the same line in the laboratory according to the
relation 

 \begin{equation}
   \lambda_{\mathrm{obs},i}  =  \lambda_{0,i} (1 + z)(1+ K_i \frac{\Delta \mu}{ \mu}),
 \end{equation}   
where $K_i$ is the sensitivity coefficient of the $i$th component computed theoretically for the Lyman and
Werner bands of the H$_2$
molecule.  Using this expression, the cosmological
redshift of a line can be distinguished from the shift due to a variation of
$\mu$.

This method was used   to obtain upper bounds on
the secular variation of the electron-to-proton mass ratio from observations of
distant absorption systems in the spectra of quasars at several redshifts.
The quasar absorption system towards QSO 0347-383 was first studied by us  using
high-resolution spectra obtained with the very large
telescope/ultraviolet-visual
echelle spectrograph (VLT\footnote{Very Large Telescope at the European Southern
Observatory
 (ESO)}/UVES\footnote{UV-Visual Echelle Spectrograph}) commissioning data we
derived a first stringent
bound
   at $(-1.8 \pm 3.8) \cdot 10^{-5}$ (Levshakov et al. \cite{Levshakov02}).
 Subsequent   measures  of  the quasar absorption systems of  QSO 
0347-382  and QSO  1232+082 provided hints for a variation $(2.4 \pm 0.6) \cdot
10^{-5}$, i.e.
at  3.5 $\sigma$  (Reinhold et al. \cite{Reinhold06}; Ivanchik et al.
\cite{Reinhold06,Ivanchik05}; Ubachs et al. \cite{Ubachs07}). The new
analysis used additional high-resolution
spectra and updated laboratory data of the
energy levels   and of the rest frame wavelengths of the
\hhh molecule.  

However,  more recently King et al. (\cite{King08}), Wendt \& Reimers
(\cite{Wendt08}) and 
Thompson et al. (\cite{Thompson09}) 
re-evaluated data of the same system and report a result in agreement with no
variation.
The  more stringent limits on   $\Delta \mu / \mu$ have been found  at 
$\Delta\mu / \mu  = (2.6\pm 3.0_{\rm
stat})\times10^{-6}$  from the combination of three H$_2$ systems (King et al.
\cite{King08}) and 
a fourth one have provided $\Delta\mu / \mu  = (+5.6\pm 5.5_{\rm stat}\pm
2.7_{\rm sys})\times10^{-6}$
(Malec et al. \cite{Malec10}).

This work is motivated on one side  by the use of  a new data set available in
the ESO data archive and previously
overlooked and by numerous findings of different groups that partially are in
disagreement witch each other. A large part of these discrepancies reflect the
different
methods of handling systematic errors. Evidently systematics are not yet under
control
or fully understood. We try to emphasize the importance to take these errors, in
particular
calibration issues, into account and put forward some measures adapted to the
problem.

The bounds on the variation of $\mu$ are generally  obtained
by using the vibro-rotational transitions of molecular hydrogen, since H$_{2}$
is a very abundant molecule   although very rarely seen in quasar absorbers.
Only a few studies  used other molecules since 
they are difficult  to detect and measure  accurately
  at large redshifts.  In general these methods provide less stringent bounds
{\chg on $\Delta\mu$ 
directly or bear a greater danger of nonuniform absorbers.
Comparisons between the redshifts of H\,I 21cm (hyperfine) measured in the radio regime
and ultraviolet resonance dipole transitions are sensitive to
changes in $X \equiv g_p\alpha^2/\mu$ (see, i.e.,  Kanekar \cite{Kanekar2010}), which offers an
important complementary verification of measured variations, although they sample
different gas volumes.}

 One  remarkable exception is the inverse spectrum of ammonia  at radio
wavelengths. 
A variation of  $\Delta \mu/\mu$ can be tested
through precise measurements of the relative radial velocities of narrow
molecular lines observed in the cold interstellar molecular cores.
This approach is based on a new method derived by Flambaum \& Kozlov
(\cite{Flambaum07}).
Ammonia NH$_3$ is a molecule whose inversion transitions  are very
sensitive to changes in $\mu$ due to a tunneling
effect. 
The sensitivity coefficient of the inversion transition NH$_3 (1,1)$
at $\nu = 23.7$ GHz 
is almost two orders of magnitude 
more sensitive to $\mu$-variation than H$_2$ molecular rotational frequencies. 
By comparing the observed inversion
frequency of NH$_3$(1,1) with a suitable rotational frequency of another
molecule arising {\it co-spatially} with ammonia, a limit on the spatial
variation of $\mu$ can be determined.

Ammonia has been detected in absorption in the 
 main gravitational lenses of the quasars B 0218+357 and PKS 1830-211. Flambaum
\& Kozlov (\cite{Flambaum07})
 combine the three detected NH$_3$  absorption spectra from
B0218+357 with rotational spectra of CO, HCO$^+$, and HCN to place a limit of
(0.6$\pm$ 1.9) $\cdot$ 10$^{-6}$  for a look-back time of 6  Gyr (redshift z =
0.68).
Accounting in detail for the velocity structure of the line profiles, Murphy et
al.
(2008) reanalyzed the ammonia data in combination with newly obtained high
signal-to-noise rotational spectra of HCO$^+$ and HCN. This yields  $<$ 1.8
$\cdot$
10$^{-6}$ at a 95\% confidence level. 
Analyzing the ten NH$_3$ inversion lines and a similar number of rotational
transitions from other molecules
Henkel et al. (2009) obtain 10$^{-6}$  as a firm upper
limit for a look-back time of 7 Gyr (z=0.89). 
However,  the  low number of NH$_3$ sources limit this method considerably, in
particular for high redshifts.

{\chg Radio observations of OH lines at the Arecibo Telescope and the Westerbork Synthesis Radio Telescope yield 
$\Delta G/G = (-1.18 \pm 0.46) \cdot 10^{-5}$ for a look back time of 2.9 Gyrs.
$G$ is defined as $G \equiv g_p [\mu\alpha^2]^{1.85}$ and the results correspond to a change
 in $\alpha$, $\mu$ and/or $g_p$ (Kanekar et al. \cite {Kanekar2010b}b). 
}

In the following we will concentrate on the  single \hhh system observed towards
QSO 0347-383 to trace the
proton-to-electron mass ration $\mu$
at high redshift  ($z_{\mathrm{abs}}=3.025$). We intend to reach a robust
estimation of the achievable
accuracy with current data by comparing independent observation runs.

\section{Data}
\subsection{Observations}

QSO 0347-383 is a bright quasar (V $\approx$ 17.3) with z$_{em}$
3.23, which shows a
Damped Lyman $\alpha$  system at z$_{abs}$ = 3.0245. The  hydrogen column
density  
is of N(H I)= $5 \cdot 10^{20}$  cm$^{-2}$ with a  rich
absorption-line spectrum  (Levshakov et al. \cite{Levshakov02}). The z$_{abs}$ = 3.025 DLA
exhibits a multicomponent velocity
structure. There are at
least two gas components: a warm gas seen in lines of
neutral atoms, H and low ions, and a hot gas where the 
resonance doublets of C IV and Si IV are formed. In correspondence of the cool
component 
molecular hydrogen was first detected by Levshakov et al (\cite{Levshakov02b}) who identified 88
H$_2$ lines.

All works on QSO 0347-383 are based on the same UVES VLT
observations\footnote{Program ID 68.A-0106.} in January 2002 (see Ivanchik
et al. \cite{Ivanchik05}).
The data used therein were retrieved from the VLT archive along with the
MIDAS based UVES pipline 
reduction procedures. The slit width was 0.8\arcsec. The grating angle for the
QSO 0347-383 observations had
a central wavelength of 4300 \AA. The images are 2$\times $2 binned. The 9
spectra were recorded during 3 nights with
an exposure time of 4500 seconds each. Additional observational parameters are
described in Ivanchick et al. (\cite{Ivanchik05}).
The above mentioned data was recently carefully reduced again by Thompson et al.
(\cite{Thompson09}). 
This work, however, is only in part based on this  original reduced data.

Here we take into account  additional observational data of QSO 0347-383  
acquired
in 2002 at the same
telescope but  not previously analyzed   \footnote{Program ID 68.B-0115(A).}.

The UVES observations comprised of 6 $\times$ 80 minutes-exposures of QSO
0347-383
on
several nights, thus adding another 28.800 seconds of exposure time. The journal of
these observations as well as additional
information is reported in Table \ref{table:1}.
Three UVES spectra were taken with the DIC1 and setting 390+580 nm and three
spectra with 
DIC2  and setting 437+860, thus providing   blue spectral ranges between 320-450
and 373-500 nm respectively.
We note that  QSO  0347-383 has no flux below 370 nm due to the Lyman
discontinuity of the z$_{\mathrm{abs}}$=3.023
absorption system. The slit width was
set to 1\arcsec\, for all observations providing a Resolving Power of $\approx$
40\,000. {\chg The different slit widths and hence different resolutions of the
observation runs pose no problem since all data are analyzed separately during
the fit.}
The seeing was varying in the range between  0.5\arcsec\ to 1.4\arcsec\ as
measured by DIMM but generally is
better than this at the telescope. 
The CCD pixels were binned   by 2$\times$2 providing an effective 0.027-0.030
\AA\ pixel, or 2.25 \kms at 400 nm
along dispersion direction.

\begin{table*}
\caption{Journal of the observations}             
\label{table:1}
\label{observdates}      
\centering          
\begin{tabular}{c c c c l l l }     
\hline\hline       
Date  & Time  & $\lambda$ &Exp(sec) & Seeing (arcsec)  & airmass &  S/N (mean) 
\\ 
\hline                    
2002-01-13& 03:42:54   & 390 &4800  &   1.7  &  1.5 &  20\\
2002-01-14 & 02:13:24 & 390 &4800    &   1.0 &  1.2 &  28 \\
 2002-01-15& 00:43:32  & 437  &4800  &  0.96   & 1.0  & 67   \\
 2002-01-18 & 03:25:04  & 437 & 4800 &   1.63  & 1.4   & 49  \\
2002-01-24 & 02:20:14  &437& 4800 &     1.07 & 1.7   & 29 \\  
 2002-02-02& 01:33:58  & 390 & 4800  &  0.5  & 1.2  &  37 \\  
\hline                  
\end{tabular}
\end{table*}

\subsection{Reduction}

The standard UVES pipeline has been followed for the data reduction. 
This includes sky subtraction and optimal extraction of the spectrum. Typical
residuals of the wavelength calibrations were of $\approx$ 0.5 \ma\, or
$\approx$ 
40 \ms\ at 400 nm. The spectra were reduced to barycentric coordinates and air
wavelengths  have
been transformed to vacuum by means of  the dispersion formula by  Edlen
(\cite{Edlen66}).
Proper calibration and data reduction will be the key to detailed analysis of
potential variation of fundamental constants. The influence of calibration
issues on the data quality is hard to measure and the magnitude of the resulting
systematic error is under discussion.
The measurements rely on detecting a pattern of small relative wavelength shifts
between different
transitions spread throughout the spectrum. 
Normally, quasar spectra are calibrated by comparison with spectra
of a hollow cathode thorium lamp  rich in unresolved spectral lines.  However
several factors are affecting the quality of the wavelength
scale. The paths for ThAr light and quasar light through the spectrograph are
not
identical  thus introducing  small distortions between ThAr and quasar
wavelength scales.    
In particular differences in the slit illuminations are not traced by the
calibration lamp. Since  source centering into the slit is varying from one
exposure to another this induce an offset in the zero point of the scales of
different frames which could be up to few hundred of \ms. In section 3.1 we
provide an estimate of these offsets which result of a mean offset of 168 m
s$^{-1}$ as well as a procedure to avoid this problem.
Laboratory wavelengths are know with  limited precision which is
varying from line to line from about 15 \ms of the better known lines to more
than 100 \ms for the more poorly known lines (Murphy et al. \cite{Murphy08}
and Thompson et al. \cite{Thompson09}).  However, this is the error which is
reflected in the size of the residuals of the wavelength calibration.
{\chg Iodine cell based calibration cannot be applied directly in the case of QSO 0347-383 since at a redshift of $z\approx3$ all observed lines lie outside the range of Iodine lines which cover about 5\,000-6\,000 \AA. Additionally, at the given level of continuum contamination due to Lyman-$\alpha$ forest at such redshifts a super-imposed spectrum of the iodine cell is not desirable.}

Effects of this kind have been investigated  at the 
Keck/HIRES spectrograph by comparing the ThAr wavelength scale with one
established from I2-cell observations
of a bright quasar by  Griest et al. (\cite{Griest10}). They found both 
absolute and relative wavelength
offsets in the Keck data reduction pipeline which can be as large as  500 - 1\,000
\ms 
 for the observed wavelength range.  Such errors would correspond  to
$\Delta\mathrm{\lambda}\approx 10-20\, \ma$ and exceed by one order of magnitude
presently quoted errors (Thompson et al. \cite{Thompson09}).
Examination of the UVES spectrograph at the 
VLT carried out via solar spectra reflected on asteroids with known radial
velocity  showed no such dramatic offsets being less than $\approx$  100
\ms  ( Molaro et al.\cite{Molaro08b}) but systematic errors at the level of few hundred \ms have
been  revealed also in
the UVES data
by comparison of relative shifts of lines with comparable response  to changes
of fundamental constants (Centurion et al. \cite{Centurion09}). These examples
well show that 
current $\Delta\mu/\mu$-analysis based on quasar absorption spectra at the level
of a few ppm enters the regime
of calibration induced systematic errors. While awaiting a new generation of
laser-comb-frequency calibration, today's efforts to investigate
potential variation of fundamental physical constants require factual
consideration of the strong systematics.

 We note also that the additional observations considered here were taken for
other purposes and the ThAr lamps are taken
during daytime,  which means several hours before the science exposures and
likely under different thermal and pressure conditions. However,
in the present work we  bypass the possibility of different zero points of the
different images via the seldom case of independent observations. Instead of
co-adding all the spectra we compute first the global velocity shifts between
the spectra with the procedure described in the following section and  we also
utilize the whole uncertainties coming from the wavelength accuracies as part of
the analysis procedure.
\begin{figure}[]
\resizebox{\hsize}{!}{\includegraphics[clip=true]{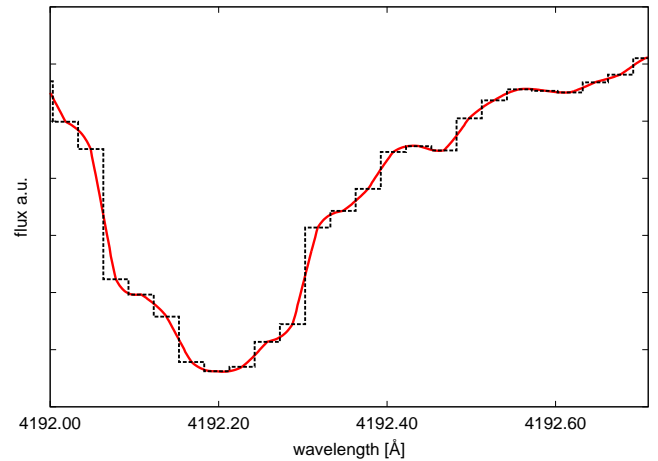}}
\caption{\footnotesize
{\chg Example region of the QSO 0347-383 spectrum showing the recorded flux ({\it dashed}) and
its interpolation via a } polynomial using {\it Neville's
algorithm} to conserve the local flux.}
\label{sampled}
\end{figure}
\begin{figure}[]
\resizebox{\hsize}{!}{\includegraphics[clip=true]{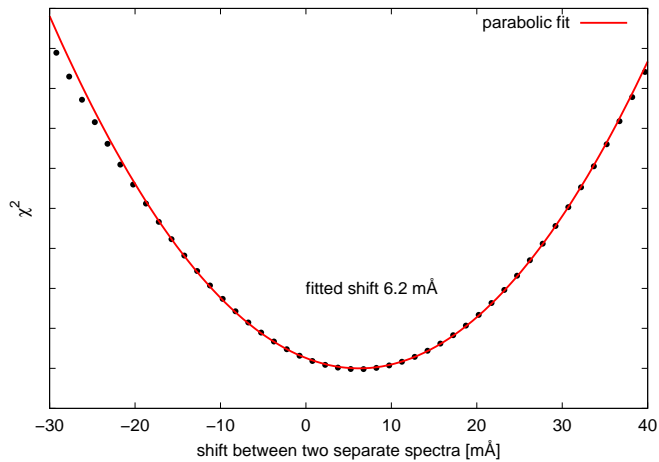}}
\caption{\footnotesize
Exemplary plot of the sub-pixel cross-correlation. The resulting shift is
ascertained
via parabolic fit. In this case the two spectra are in best agreement with a
relative shift of 6.2 $\pm$ 0.5 \ma.}
\label{eta}
\end{figure}
\subsection{Noise level}
The UVES data reduction procedure deliver the error spectrum along the optimally
 extracted spectrum.
The given error in flux of all 15 spectra was tested against the zero level
noise in saturated areas. A broad region of saturated absorption is available
near $3906 \AA$ in the
observers frame. Statistical analysis revealed a variance corresponding to
$\sim$120\% of the given
error on average for the 15 spectra.  This means that normally errors that rely
to the standard extracted routine
are probably  underestimated by a comparable amount.
In particular we compared the standard deviation of the flux between 
$3903.8 \AA$ and $3908.7 \AA$ (roughly 160 samples) with the average of the
specified error for that range.
In our analysis for each of the spectra the calculated correction factor was
applied.
\section{Preprocessing}
\subsection{Relative shifts of the 15 spectra}

Prior to further data processing
the reduced spectra are reviewed in detail.
The first data set (henceforward referred to as set A) consists of nine separate
spectra observed between 7th and 9th of January in 2002. The second set of 6
spectra (B) was obtained between January 13th and February 2nd in 2002 (see
Table \ref{observdates}). 

\label{preprocessing}

Due to slit illumination effects  and grating motions the individual
spectra are subject to small shifts -- commonly on sub-pixel level -- in
wavelength. These shifts will be particularly crucial in the process of
co-addition of several exposures.
To estimate these shifts  all spectra were interpolated by a polynomial using
{\it Neville's algorithm} to conserve the local flux (see Fig. \ref{sampled}).
The resulting pixel step on average is $1/20$ of the original data. Each
spectrum was compared to the others. For every data point in a spectrum the
pixel with the closest wavelength was taken from a second spectrum. Their
deviation in flux was divided by the quadratic mean of their given errors in
flux. This procedure was carried out for all pixels inside certain selected
wavelength intervals resulting in a mean
deviation of two spectra.
The second spectrum is then shifted against the first
one in steps of $\sim1.5 $ \ma. The run of the discrepancy of two spectra is
of parabolic nature with a minimum at the relative shift with the best
agreement. Fig. \ref{eta} shows the resulting curve with a parabolic fit. In
this exemplary case the second spectrum shows a shift of 6.2 
\ma\ in relation to
the reference spectrum.
The clean parabolic shape verifies the approach.  Table \ref{shifts} shows the
corresponding offsets for the 15 spectra.
 The offsets between the exposures are relevant with a peak to peak excursion up
to almost 800 \ms. The average deviation is 2.3 \ma or 170 \ms at
400 nm.
For further analysis in this paper all the 15 spectra  are shifted to their
common mean, which is taken as an arbitrary reference position.
{\chg To avoid areas heavily influenced by cosmic
events  or areas close to
overlapping orders only certain wavelength intervals were taken into account,
namely the regions 3877-3886 \AA, 3986-4027 \AA, 4145-4175 \AA \,and 4216-4240
\AA\, (referred to as 1-4 in Table \ref{shifts}). The individual intervals show no significant differential shifts and their combined
wavelength range was used to obtain a more robust mean shift between two spectra. To estimate the error of the given shifts, the routine was carried out for all spectrum-combinations and the resulting shifted spectra were checked for shifts again. We estimate the error per shift to be of 38 \ms based on the deviation of the individual shifts of four wavelength intervals.}
 Section \ref{influence}
illustrates its influence on the data analysis with respect to the previous
analysis of the data set A, in which this effect was not considered.
\begin{table}
\caption{Relative shifts of the observed spectra to their common mean. Spectra
A1-A9 correspond
 to the observations of Program ID 68.A-0106, spectra B1-B6 to Program ID
68.B-0115(A), respectively. {\it shift 1-4} represent the individual offsets for the four selected wavelength intervals. }
\label{shifts}
\centering          
\begin{tabular}{c r r r r r r }   
\hline\hline       
ID  & shift 1& shift 2&shift 3&shift 4&shift 1-4&$\sigma_{1-4}$\\ 
	& (\ms) & (\ms) & (\ms)& (\ms)  & (\ms)& (\ms)\\
\hline
A1      &-243   &-239   &-240   &-269   &-203   &13\\
A2      &-163   &-112   &-52    &-107   &-135   &39\\
A3      &42     &92     &95     &105    &116    &24\\
A4      &-46    &-107   &-29    &-45    &-61    &30\\
A5      &222    &295    &182    &223    &268    &41\\
A6      &15     &-75    &-101   &11     &-31    &51\\
A7      &-189   &-312   &-212   &-258   &-249   &47\\
A8      &129    &11     &87     &112    &65     &45\\
A9      &305    &192    &218    &229    &249    &42\\
\hline
B1      &114    &70     &85     &30     &84     &30\\
B2      &524    &439    &563    &501    &496    &45\\
B3      &19     &17     &50     &86     &39     &28\\
B4      &-286   &-417   &-290   &-388   &-339   &58\\
B5      &129    &11     &24     &-23    &30     &57\\
B6      &-101   &-120   &-173   &-111   &-158   &28\\
\hline          
\multicolumn{6}{r}{average standard deviation}&38\\
\hline          
\end{tabular}
\end{table}

\subsection{Selection of lines and line fitting}
The selection of suitable \hhh features for the final analysis is rather 
subjective.
As a matter of course all research groups cross-checked their choice of lines
for
unresolved blends or saturation effects. The decision whether a line was
excluded due to continuum contamination or not, however, relied mainly on the
expert knowledge of the researcher and was only partially reconfirmed by the
ascertained uncertainty of the final fitting procedure.
This work puts forward a more generic approach adapted to the fact that we have
two distinct observations of the same object.

In this paper each \hhh signature is fitted with a single component. The
surrounding flux is fit by a ploynomial and the continuum is rectified
accordingly. This approach is tested and verified in Wendt \& Reimers
(\cite{Wendt08}), however lines near saturated or steeply descending areas
should be avoided. 
A selection of 52 (in comparison with 68 lines for that system by
King et al. (\cite{King08})) lines is fitted separately for each dataset
of 9 (A) and 6 (B) exposures, respectively. In this selection merely blends
readily identifiable or emerging from equivalent width analysis are
excluded.
Each rotational level is fitted with conjoined line parameters except for
redshift naturally. {\chg A common column density $N$ and broadening parameter $b$
 corresponds to each of the observed rotational levels (in this case $J=1,2,3$).}

\begin{figure}[]
\resizebox{\hsize}{!}{\includegraphics[clip=true]{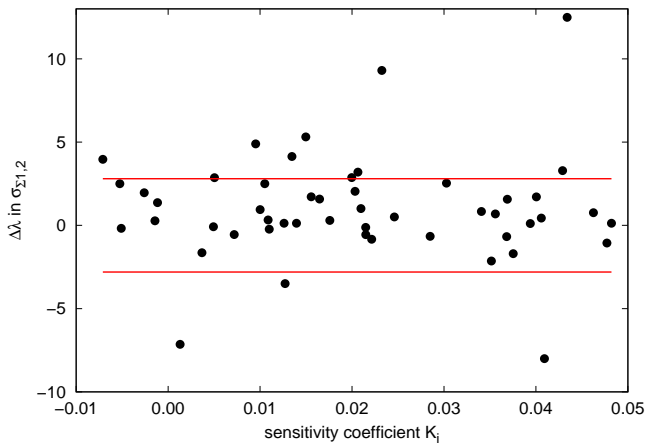}}
\caption{
\footnotesize
Selection of 52 apparently reasonable lines to be fitted separately for each
dataset
of 9 and 6 exposures, respectively. Their absolute offset
$\Delta\lambda_{\mathrm{effective}}$ to each other is expressed in relation to
their combined error given by the fit (see Eq.\ref{eq1}). The dashed lines
border the $3 \sigma$ domain.
}
\label{compare}
\end{figure}

\begin{figure}[]
\resizebox{\hsize}{!}{\includegraphics[clip=true]{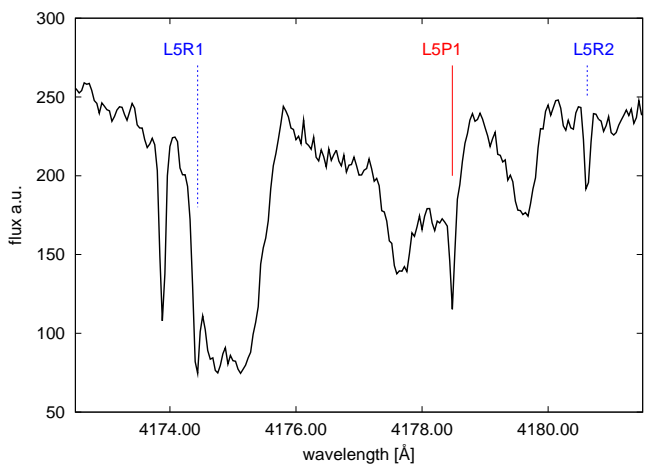}}
\caption{
\footnotesize
Part of the co added spectrum near 4176\AA. The data however, were not co added
for the fit.
L5R1 and L5R2 match the 3-$\sigma$ criterion, L5P1 does not.}
\label{3s}
\end{figure}

The data are not co-added but analyzed simultaneously via
the fitting procedure introduced in Quast et al. (\cite{Quast05}).
{\chg In principle each set of parameters (line centroid, broadening parameter, column density, coefficients of the continum polynomial) drawn from a large parameter space is tested in all individual spectra and judged by a weighted $\chi^2$.
The standard deviations of line positioning are provided by the diagonal elements of the
scaled covariance matrix, a procedure described in detail in the above mentioned paper by Quast et al. and verified, i.e., in Wendt et al. (\cite{Wendt08})}.
For each of the 52 lines there are two resulting fitted redshifts or observed
wavelengths, respectively, with their error estimates.
To avoid false confidence, the single lines are not judged by their error
estimate but by their difference in wavelength between the two data sets in
relation to the combined 
error estimate. Fig. \ref{compare} shows this dependency.
The absolute offset $\Delta\mathrm{\lambda_{\mathrm{effective}}}$ to each other
is expressed in 
relation to their combined error given by the fit:
\begin{equation}
\Delta\lambda_{\sigma_{\Sigma 1,2}} =
\frac{\Delta\lambda_{\mathrm{effective}}}{\sqrt{\sigma^2_{\lambda_1}+\sigma_{
\lambda_2}^2}}.
\label{eq1}
\end{equation}
Fig. \ref{compare} reveals notable discrepancies between the two datasets, the
disagreement is partially
exceeding the 5-$\sigma$ level\footnote{Lines fitted with seemingly high
precision and thus a low error
reach higher offsets than lines with a larger estimated error at the same
discrepancy in
 $\lambda_{\mathrm{obs}}$. Clearly the lower error estimates merely reflects the
statistical
quality of the fit, not the true value of the specific line position.}.
Since the fitting routine is known to provide proper error estimates
(Quast et al. \cite{Quast05}; Wendt et al. \cite{Wendt08}), the dominating
source of error in the determination
of line positions is due to systematic errors. This result indicates calibration
issues
of some significance at this level of precision. The comparison of two
independent observation runs
reveals a source of error that cannot be estimated by the statistical quality
of the fit alone.
For further analysis only lines that differ by less than 3 $\sigma$ are taken
into account.
This criterion is met by 36 lines. Fig. \ref{3s} shows three examplary \hhh
features corresponding to the 
transitions L5R1, L5P1, L5R2. All have similar sensitivity towards changes in
$\mu$. L5P1 fails the
applied self consistency check between the two data sets and is excluded in the
further analysis.
Table \ref{exlines} lists the excluded lines.

It is noteworthy that line selections of this absorption system by other groups
diverge from each other
by a large amount. King et al. (\cite{King08}) processed a total of 68 lines. By
reconstructing the 
continuum flux with additionally fitted lines of atomic hydrogen they felt
confident not to care about
the relative position of the \hhh features next to the Lyman-$\alpha$ forrest.
Fitting \hhh features as single lines, however, is affected by the surrounding
flux and its nature as simulations
have shown (Wendt \& Reimers \cite{Wendt08}). 
Thompson et al. (\cite{Thompson09}) selected 36 lines for analysis of which
differs from our
semi-automatical choice of lines by almost 40\%. 

\begin{figure}[]
\resizebox{\hsize}{!}{\includegraphics[clip=true]{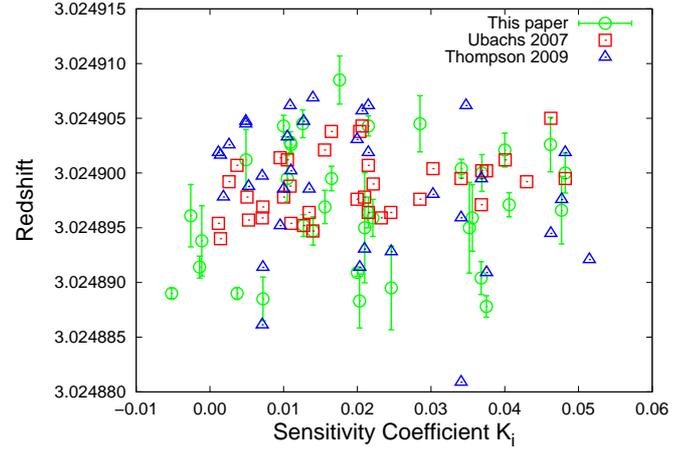}}
\caption{
\footnotesize
Final results in redshift vs. sensitivity coefficient $K_i$ for this paper
(\textit{circles}), Ubachs et
al. \cite{Ubachs07} (\textit{squares}) and Thompson et al. \cite{Thompson09}
(\textit{triangles}).
}
\label{figcompare}
\end{figure}

\begin{figure}[]
\resizebox{\hsize}{!}{\includegraphics[clip=true]{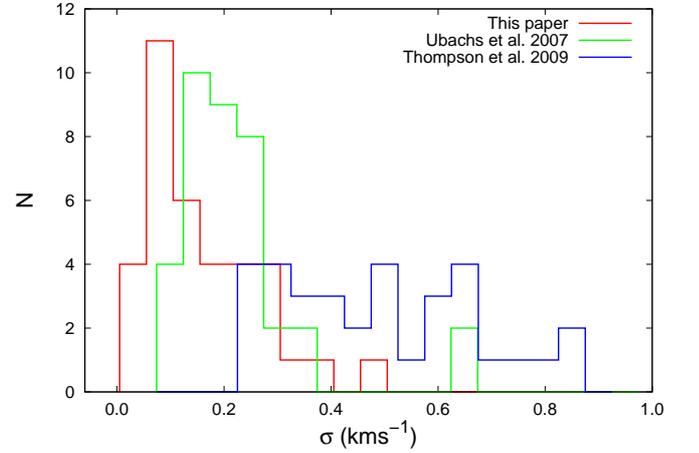}}
\caption{
\footnotesize
Line positioning errors in \kms\, for this paper (\textit{solid}), Ubachs et
al. \cite{Ubachs07} (\textit{dashed}) and Thompson et al. \cite{Thompson09}
(\textit{dotted}), binned to 50 \ms.
}
\label{figerrv}
\end{figure}

Hence, the different findings arise through in large parts independent analysis.
Different approaches, line selections and in the end applied methods contribute
to a more solid
constraint on variation of fundamental constants. This variety is mandatory to
understand contradicting findings,
not only in case of the proton-to-electron mass ratio. Table \ref{linelist}
report the molecular line position and relative errors.
Figure \ref{figcompare}
plots the results of this paper, Ubachs et al. (\cite{Ubachs07})  and Thompson et
al. (\cite{Thompson09}), who published the individual fit parameters. 
The redshifts derived are of 
   3.0248969 (56),    3.0248988(29)    and                             
3.0248987(61) for this paper, Ubachs et al (2007) and Thompson (2009)
respectively, which is not surprising being based at least partially on the same
data. 
  All three
analysis are based on the same source for sensitivity coefficients. The
distribution of positioning errors for the mentioned works is illustrated in
Figure \ref{figerrv}.  The three sets of measure show a significant scatter
around the mean quite in excess of the error in line position which is
suggestive of the presence of systematic errors.

The chosen $\Delta\lambda$ criterium for line selection permits evaluation of
the self-consistency of a 
line positioning via fit for the involved data. While the availability of two
independent observations on short
time scale is rather special, it illustrates one applicable modality to avoid
relying on the fitting 
apparatus alone.

\section{Results}
\begin{figure}[]
\resizebox{\hsize}{!}{\includegraphics[clip=true]{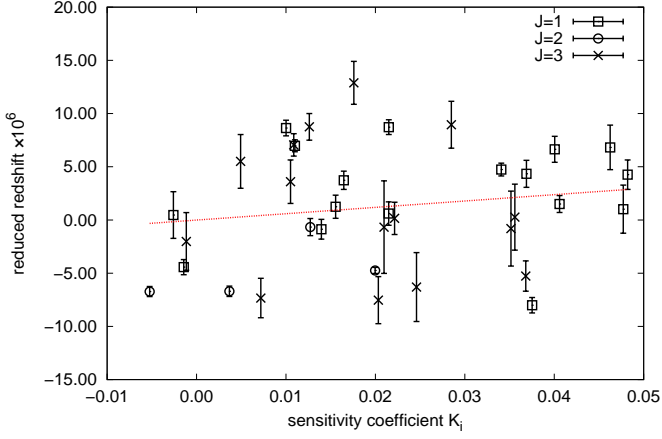}}
\caption{
\footnotesize
The unweighted fit for QSO 0347-383 corresponds to $\Delta\mu/\mu = (15 \pm 16)
\times
10^{-6}$. The error bars correspond merely to the fitting uncertainty in the
order of 180 \ms\, on average.
Note, that at such a high scatter $z_{Ki=0}$ differs from $\bar{z}$ by less than
1 $\sigma_\mathrm{z}$.}
\label{result}
\end{figure}

For the final analysis the selected 36 lines are fitted in all 15 shifted,
error-scaled 
spectra simultaneously. The result of an unweighted linear fit corresponds to
 $\Delta\mu/\mu = (15 \pm 16) \times 10^{-6}$ over the look-back time of
$\sim11.5$ Gyr for 
 $z_{\mathrm{abs}}=3.025$. Fig. \ref{result} shows the resulting plot. 
The complete list of lines is shown in Table \ref{linelist}.
The approach to apply an unweighted fit is a consequence of the unknown nature
of the
prominent systematics. The graphed scatter in redshift can not be
explained by 
the given positioning errors alone. The likeliness of the data with the
attributed error
being linearly correlated is practically zero. The fit to the data is not
self-consistent.
For this work the calibration errors and the influence of unresolved blends are
assumed to be
dominant in comparison to individual fitting uncertainties per feature.
For the following analysis the same error is adopted for each line.
With an uncertainty in redshift of $1\times 10^{-6}$ we obtain:
$\Delta\mu/\mu = (15 \pm 6) \times 10^{-6}$.
However the goodness-of-fit is below 1 ppm and is not self consistent.
{\chg The goodness of fit of a statistical model describes how well it fits a set of observations. A linear model with the given parameters does not represent the observed data sample very well. The apparent discrepancies between model and data including their errorbars shown in Figure \ref{result} are extremely unlikely  (below one part per million) to be chance fluctuations.}
Judging by that and Fig. \ref{result}, a reasonable error in observed redshift
should
at least be in the order of $4-5\times 10^{-6}$. The weighted fit gives:
$\Delta\mu/\mu = (15 \pm 14) \times 10^{-6}$.

This approach is motivated by the goodness-of-fit test:\\
$Q(\chi^2|\nu)$ is the probability that the observed chi-square will exceed the
value $\chi^2$ by 
chance even for a correct model, $\nu$ is the number of degrees of freedom.
Given in relation to the incomplete gamma 
function:
\begin{equation}
Q(\chi^2|\nu) = \Gamma\left(\frac{\nu}{2},\frac{\chi^2}{2}\right).
\end{equation}
Assuming a gaussian error distribution, $Q$ gives a quantative measure or the
goodness-of-fit
of the model. If $Q$ ist very small for some particular data set,
then the apparent discrepancies are unlikely to be chance
fluctuations that would be expected for a gaussian error distribution. Much more probably either the model is wrong or the size of the
measurement
errors is larger than stated. However, the chi-square probability $Q$ does not
directly measure the credibility of the assumption that the measurement errors
are
normally distributed. In general, models with $Q < 0.001$ can be considered
inacceptable.
In this case the model is given and hence the low probability is due to
underestimated errors in the data.
Solely for given errors of $\approx$ 300 \ms, corresponding to $\approx
4\times10^{-6}$ in redshift
for QSO 0347-383 the goodness-of-fit parameter $Q$ exceeds $0.001$.
The scale of the error appears to be $\approx$ 300 \ms to achieve a
self-consistent
fit to the data. 

\begin{figure}[]
\resizebox{\hsize}{!}{\includegraphics[clip=true]{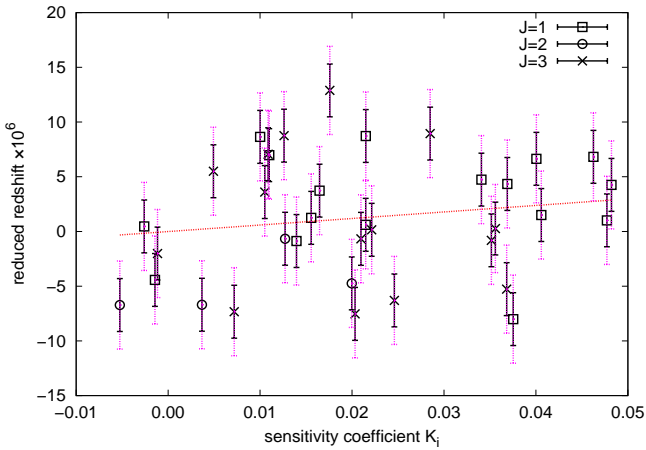}}
\caption{
\footnotesize
Data identical to Fig.\ref{result}. The errorbars represent the average
positioning error (solid)
and the additional systematic error (dotted) of $\approx$ 180 \ms\, and
$\approx$ 120 \ms, respectively.}
\label{resulterr}
\end{figure}

For the data on QSO 0347-383 this corresponds to an error in the observed
wavelength of
roughly
$4 \ma$, which is notably larger than the estimated errors for the individual
line fits which 
ranges from $0.5 \ma$\, to $6.5 \ma$\, with an average of $2.5 \ma$ ($\approx$
180 \ms).
The systematic error contributes an uncertainty of about $2\ma$ on
average. The immediate calibration errors are in the order of 50 \ms for set B
and presumably slightly larger for set A.
Figure \ref{resulterr} plots the data with errorbars corresponding to 180\ms\, 
 and the total of 300 \ms.

The final result can be subdivided as:\newline
$\Delta\mu/\mu = \left(15 \pm (9_{\mathrm{stat}} + 6_{\mathrm{sys}})\right) \times
10^{-6}$.

The comparably high scatter in Figure \ref{result} can partially be attributed
to the approach to
fit single \hhh components with a polynomial fit to the continuum.
In special cases, contaminated flux bordering a \hhh signature can introduce
additional uncertainty
in positioning  therefore checks for self-consistency and systematics are of
utmost importance.

The determination of the different errors involved is on a par with the actual
result.

We believe that this result represents the limit of accuracy that can be reached
with the given data set and
 the applied methods for analysis. The presented method yields a null result.
The recent work by Thompson et al. (\cite{Thompson09}) stated 
$\Delta\mu/\mu = (-28 \pm 16) \times 10^{-6}$ for a weighted fit based on the
same system in QSO 0347-383. The therein stated error reflects
 the statistical uncertainty alone.

Note, that the given systematics of $2.7$ppm for Keck/HIRES data given in Malec
et al. (\cite{Malec10}) are in first approximation estimated by the observed
$\sim$500\,\ms\
peak-to-peak intra-order value reduced according to the number of
molecular transitions observed, i.e.~$\sim500\,\ms/\sqrt{93} \approx
52\,\ms$. 

This work proposes to take alternative approaches into account when operating
that close
to the limits of several involved systems. The presented method was applied to
the known QSO 0347-383 data
for two main reasons:
\begin{itemize}
\item As an example to put forward alternative strategies.
\item A stand-alone determination of $\Delta\mu/\mu$ based on QSO 0347-383 to
back up current null-results
and consequential constraints.
\end{itemize}

\begin{table*}
\caption{QSO 0347-383 Line List}
\label{linelist}
\centering          
\begin{tabular}{l r r r r r r }     
\hline\hline       
Line ID & K factor & Obs. wavelength [\AA] & pos. error [\AA] & Lab. wavelength
[\AA] & pos. error [\kms] & redhisft\\ 
\hline
L14R1 & 0.0462 & 3811.5038 & 0.0031 & 946.9804 & 0.247 & 3.0249025 \\
W3Q1 & 0.0215 & 3813.2825 & 0.0012 & 947.4219 & 0.091 & 3.0249043 \\
W3P3 & 0.0210 & 3830.3795 & 0.0064 & 951.6719 & 0.499 & 3.0248950 \\
L13R1 & 0.0482 & 3844.0442 & 0.0023 & 955.0658 & 0.181 & 3.0248999 \\
L13P1 & 0.0477 & 3846.6271 & 0.0039 & 955.7083 & 0.306 & 3.0248966 \\
W2Q1 & 0.0140 & 3888.4352 & 0.0017 & 966.0961 & 0.128 & 3.0248948 \\
W2Q2 & 0.0127 & 3893.2050 & 0.0013 & 967.2811 & 0.099 & 3.0248951 \\
L12R3 & 0.0368 & 3894.7939 & 0.0019 & 967.6770 & 0.149 & 3.0248904 \\
W2Q3 & 0.0109 & 3900.3288 & 0.0013 & 969.0492 & 0.097 & 3.0249028 \\
L10R1 & 0.0406 & 3952.7477 & 0.0015 & 982.0742 & 0.110 & 3.0248972 \\
L10P1 & 0.0400 & 3955.8160 & 0.0020 & 982.8353 & 0.154 & 3.0249022 \\
L10R3 & 0.0356 & 3968.3977 & 0.0040 & 985.9628 & 0.302 & 3.0248960 \\
L10P3 & 0.0352 & 3975.6657 & 0.0055 & 987.7688 & 0.412 & 3.0248950 \\
W1Q2 & 0.0037 & 3976.4877 & 0.0007 & 987.9745 & 0.054 & 3.0248890 \\
L9R1 & 0.0375 & 3992.7546 & 0.0013 & 992.0164 & 0.098 & 3.0248877 \\
L9P1 & 0.0369 & 3995.9594 & 0.0022 & 992.8096 & 0.167 & 3.0249000 \\
L8R1 & 0.0341 & 4034.7699 & 0.0011 & 1002.4521 & 0.085 & 3.0249004 \\
L8P3 & 0.0285 & 4058.6575 & 0.0034 & 1008.3860 & 0.255 & 3.0249046 \\
W0R2 & -0.0052 & 4061.2132 & 0.0006 & 1009.0249 & 0.047 & 3.0248890 \\
L7P3 & 0.0246 & 4103.3836 & 0.0052 & 1019.5022 & 0.379 & 3.0248894 \\
L6R3 & 0.0221 & 4141.5640 & 0.0023 & 1028.9866 & 0.168 & 3.0248960 \\
L6P3 & 0.0203 & 4150.4349 & 0.0034 & 1031.1926 & 0.245 & 3.0248882 \\
L5R1 & 0.0215 & 4174.4204 & 0.0019 & 1037.1498 & 0.139 & 3.0248963 \\
L5R2 & 0.0200 & 4180.6152 & 0.0005 & 1038.6903 & 0.034 & 3.0248910 \\
L5R3 & 0.0176 & 4190.5690 & 0.0031 & 1041.1588 & 0.218 & 3.0249086 \\
L4R1 & 0.0165 & 4225.9822 & 0.0016 & 1049.9597 & 0.111 & 3.0248994 \\
L4P1 & 0.0156 & 4230.2974 & 0.0021 & 1051.0325 & 0.151 & 3.0248969 \\
L4R3 & 0.0126 & 4242.1531 & 0.0018 & 1053.9761 & 0.126 & 3.0249045 \\
L4P3 & 0.0105 & 4252.1911 & 0.0030 & 1056.4714 & 0.211 & 3.0248994 \\
L3R1 & 0.0110 & 4280.3234 & 0.0010 & 1063.4601 & 0.071 & 3.0249027 \\
L3P1 & 0.0100 & 4284.9349 & 0.0014 & 1064.6054 & 0.097 & 3.0249043 \\
L3R3 & 0.0072 & 4296.4822 & 0.0028 & 1067.4786 & 0.198 & 3.0248884 \\
L3P3 & 0.0049 & 4307.2114 & 0.0040 & 1070.1409 & 0.276 & 3.0249012 \\
L2P3 & -0.0011 & 4365.2399 & 0.0046 & 1084.5603 & 0.318 & 3.0248937 \\
L1R1 & -0.0014 & 4398.1291 & 0.0015 & 1092.7324 & 0.100 & 3.0248913 \\
L1P1 & -0.0026 & 4403.4456 & 0.0042 & 1094.0520 & 0.283 & 3.0248961 \\
\hline          
average & & & 0.0025 & & 0.184& \\
\end{tabular}
\end{table*}
\begin{table}
\caption{Excluded lines}
\label{exlines}
\centering          
\begin{tabular}{l r r r r r }     
\hline\hline       
Line ID & K factor & Lab. wavelength [$\AA$]\\
\hline
L12P2 & 0.0434 & 966.2755 \\
L11P1 & 0.0429 & 973.3345 \\
L11P2 & 0.0409 & 975.3458 \\
W0Q2 & -0.0071 & 1010.9384 \\
L7R1 & 0.0303 & 1013.4370 \\
L6R2 & 0.0245 & 1026.5283 \\
L6P2 & 0.0232 & 1028.1058 \\
L5P1 & 0.0206 & 1038.1571 \\
L5P2 & 0.0186 & 1040.3672 \\
L4R2 & 0.0150 & 1051.4985 \\
L4P2 & 0.0135 & 1053.2843 \\
L3R2 & 0.0095 & 1064.9948 \\
L2R1 & 0.0050 & 1077.6989 \\
L2R3 & 0.0013 & 1081.7113 \\
\end{tabular}
\end{table}
\subsection{Uncertainties in the sensitivity coefficients}
At the current level of precision, the influence of uncertainties in the
sensitivity 
coefficients $K_i$ is minimal. It will be of increasing importance though when
wavelength calibration can be improved by pedantic demands on future
observations. Eventually Laser Frequency Comb calibration will allow for
practically arbitrary
precision
and uncertainties in the calculations  of sensitivities will play a role.
Commonly the weighted fits neglects the error in $K_i$.

Effective analysis in the future involves consideration of the error budget of
the sensitivity coefficients.
The $\chi^2$ merit function for the generic case of a straight-line fit with
errors in both coordinates
is given by:
\begin{equation}
\chi^2(a,b) =
\sum^{N-1}_{i=0}\frac{(y_i-a-bx_i)^2}{\sigma^2_{yi}+b^2\sigma^2_{xi}}
\end{equation}
and can be solved numerically with valid approximations (Lybanon
\cite{Lybanon84}).

At the current level even an error in $K_i$ of about 10\% merely has an impact
on
the error estimate
in the order of a few $10^{-6}$, as resulted from simulations. The factual
errors are expected
to be in the order of merely a few percent (Ubachs et al. \cite{Ubachs07}), yet
they might contribute to the precision of
future analysis.

Alternatively the uncertainties in $K_i$ can be translated into an uncertainty
in redshift via
the previously fitted slope:
\begin{equation}
\sigma_{z_{i\,\mathrm{total}}}=\sigma_{z_{i}}+b \times \sigma_{K_{i}}
\hspace{0.5cm} \mathrm{with}\hspace{0.5cm} b=(1+z_{\mathrm{abs}}) \frac{\Delta
\mu}{\mu}.
\end{equation}
The results of this ansatz are similar to the fit with errors in both
coordinates and in general this is simpler to implement.

Another possibility is to apply a gaussian error to each sensitivity coefficient
and redo the
normal fit multiple times with alternating variations in $K_i$.
Again, the influence on the error-estimate is in the order of 1 ppm.
The different approaches to the fit allow to estimate its overall robustness as
well.
\subsection{Individual line pairs}
\begin{figure}[]
\resizebox{\hsize}{!}{\includegraphics[clip=true]{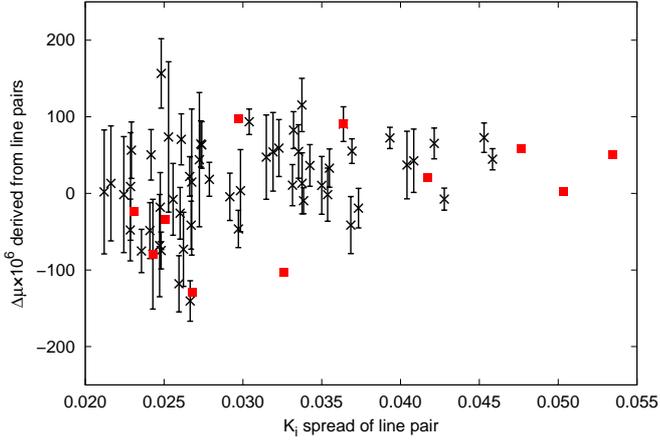}}
\caption{
\footnotesize
$\Delta\mu/\mu$ derived from individual line pairs (52) which are separated by
less
than 110 \AA\  and show a difference in sensitivity of more then $0.02$. The
errorbars reflect
the combined positioning error of the two contributing lines.
 The weighted fit corresponds to $\Delta\mu/\mu = (6 \pm
12)\times10^{-6}$.\newline
 The \textit{filled squares} graph 11 line pairs, selected to give the largest
difference in sensitivity ($\geq 0.02$) towards 
variation in $\mu$ (See Table \ref{11lines}).
}
\label{dmu}
\end{figure}

 $\Delta\mu/\mu$ can also be obtained by using merely two lines that show
different sensitivity towards changes
in the proton-to-electron mass ratio. Another criterion is their separation in
the wavelength frame to
avoid pairs of lines from different ends of the spectrum and hence in particular
error-prone.
Several tests showed that a separation of $\Delta\lambda \leq 110\, \AA$ and a
range of sensitivity
coefficients $K_1-K_2 \geq 0.02$ produces stable results that do not change any
further with more stringent criteria.
Pairs that cross two neighboring orders ($\sim50 \AA$) show no striking
deviations either.
Fig. \ref{dmu} graphs the different values for $\Delta\mu/\mu$ derived from 52
line pairs that match the
aforementioned criteria. Note, that a single observed line contributes to
multiple pairs. The gain in
statistical significance by this sorting is limited as pointed out by Molaro et
al. (\cite{Molaro08}).
Their average value yields  $\Delta\mu/\mu = 6 \pm 12 \times10^{-6}$.
The scatter is then related to uncertainties in the wavelength determination
which is mostly due to calibration
errors. The standard error is $8 \times 10^{-6}$.

The approach to use each observed line only once in the analysis is plotted in
Fig. \ref{dmu}
as filled squares. The pairs to derive  $\Delta\mu/\mu$ from were constructed by
grouping the line
with the highest sensitivity value together with the line corresponding to the
lowest value for K$_i$
and so on with the remaining lines. The distance in wavelength space between the
two lines was no
critierum and it ranges from 20\AA\, to 590\AA\, (see Table \ref{11lines}).
Without reutilization of lines, 11 pairs with a coverage in sensitivity of
$\Delta$K$_i \geq 0.02$ were found.

Evidently the usage of lines with comparably large distances in the spectrum has
no influence on the results.
\begin{table}
\caption{Grouping all observed lines into 17 pairs of maximum K$_i$ sensitivity
not 
considering their separation in wavelength space (\textit{rightmost} column).}
\label{11lines}
\centering          
\begin{tabular}{l l r r r}     
\hline\hline       
Line 1 & Line 2 & $\Delta\mu/\mu$ & $\Delta$K$_i$ & $\Delta\lambda$ [$\AA$]\\
\hline
W0R2 & L13R1 & 50.5 $\times 10^{-6}$ & 0.0535 & -217.169 \\
L1P1 & L13P1 & 2.7 $\times 10^{-6}$ & 0.0503 & -556.818 \\
L1R1 & L14R1 & 58.6 $\times 10^{-6}$ & 0.0477 & -586.625 \\
L2P3 & L10R1 & 20.9 $\times 10^{-6}$ & 0.0417 & -412.492 \\
W1Q2 & L10P1 & 90.4 $\times 10^{-6}$ & 0.0364 & -20.672 \\
L3P3 & L9R1 & -103.0 $\times 10^{-6}$ & 0.0326 & -314.457 \\
L3R3 & L9P1 & 97.6 $\times 10^{-6}$ & 0.0297 & -300.523 \\
L3P1 & L12R3 & -128.9 $\times 10^{-6}$ & 0.0268 & -390.141 \\
L4P3 & L10R3 & -33.9 $\times 10^{-6}$ & 0.0251 & -283.793 \\
W2Q3 & L10P3 & -79.4 $\times 10^{-6}$ & 0.0243 & 75.337 \\
L3R1 & L8R1 & -24.0 $\times 10^{-6}$ & 0.0231 & -245.554 \\
\hline
L4R3 & L8P3 & 3.1 $\times 10^{-6}$ & 0.0159 & -183.496 \\
W2Q2 & L7P3 & -120.0 $\times 10^{-6}$ & 0.0119 & 210.179 \\
W2Q1 & L6R3 & 34.0 $\times 10^{-6}$ & 0.0082 & 253.129 \\
L4P1 & W3Q1 & 312.7 $\times 10^{-6}$ & 0.0059 & -417.015 \\
L4R1 & L5R1 & -154.6 $\times 10^{-6}$ & 0.0050 & -51.562 \\
L5R3 & W3P3 & -996.4 $\times 10^{-6}$ & 0.0034 & -360.190 \\
\end{tabular}
\end{table}

\subsection{Influence of the preprocessing} \label{influence}
\begin{figure}[]
\resizebox{\hsize}{!}{\includegraphics[clip=true]{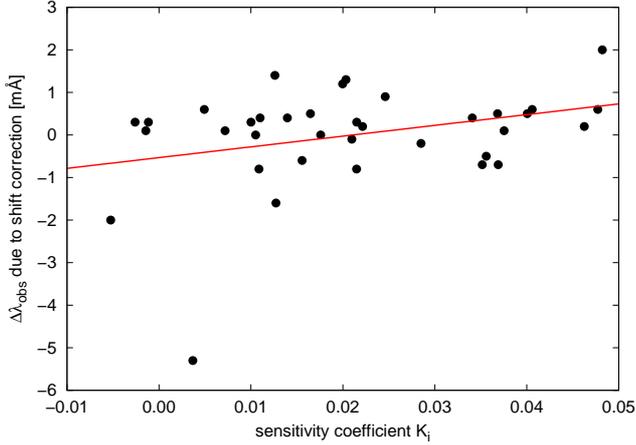}}
\caption{
\footnotesize
Variation in fitted positions for all lines with and without initial
correction for shifts in between the 15 spectra. The slope of the fit
is dominated by three lines. 
}
\label{shift_influence}
\end{figure}

Section \ref{preprocessing} describes the initial shift to a common mean of all
15 spectra.
The complete analysis was redone with error-scaled but unshifted spectra and the
ascertained line positions
of both runs compared. Fig. \ref{shift_influence} shows the difference for each
\hhh line in m\AA\ over
the corresponding sensitivity coefficients $K_i$. The plotted line is a straight
fit. Clearly the slope
is dominated by three individual lines whose fitted centroids shifted up to
$5.5$ m\AA\ due to the preprocessing.
These three lines in particular produce a trend towards variation in $\mu$ when
grating shifts and other effects
are not taken into account. This single-sided trend probably occurred by mere
chance but at such low statistics it influences the final result. Similar
effects might have
 introduced trends of non-zero variation in former works
(i.e., Ivanchik et al. \cite{Ivanchik05}; Reinhold et al. \cite{Reinhold06}).
\begin{acknowledgements}
We are thankful for the support from the Collaborative Research Centre 676 and
for
helpful discussions on this topic with D. Reimers, S.A. Levshakov, P. Petitjean
and M.G. Kozlov.
\end{acknowledgements}

\bibliographystyle{aa}

\end{document}